\documentclass{article}
\usepackage{frascatiphys}
\newcommand{\dke}{D^0\to K^-e^+\nu_e}
\newcommand{\dpie}{D^0\to \pi^-e^+\nu_e}

\newcommand{\dpkpp}{D^+\to K^-\pi^+\pi^+}

\newcommand{\vcs}{V_{cs}}
\newcommand{\vcd}{V_{cd}}
\newcommand{\vub}{V_{ub}}
\newcommand{\vus}{V_{us}}
\newcommand{\vtd}{V_{td}}

\newcommand{\vts}{V_{ts}}
\newcommand{\vcb}{V_{cb}}
\newcommand{\vtb}{V_{tb}}

\begin{document}
\title{
Status of Charm Flavor Physics
}
\author{
I. Shipsey \\
{\em Department of Physics, Purdue University,
        West Lafayette, IN 47907, U.S.A.}
}
\maketitle
\baselineskip=11.6pt
\begin{abstract}
The role of charm in testing the
Standard Model description of quark mixing and CP violation
through measurements of lifetimes, decay constants and
semileptonic form factors is reviewed. Together with Lattice QCD,
charm  has the potential this decade to maximize the sensitivity
of the entire flavor physics program to new physics. and pave the
way for understanding physics beyond the Standard Model at the LHC
in the coming decade. The status of indirect searches for physics
beyond the Standard Model through charm mixing, $CP$-violation and
rare decays is also reported.
\end{abstract}
\baselineskip=14pt
%


\section{Introduction}

Charm plays a dual role in flavor physics. First it provides
important supporting measurements for studies of $CP$-violation in
$B$ physics. These measurements test QCD technologies such as
Lattice QCD, QCD sum rules and chiral theory. The first of these
theoretical approaches is the most promising for very precise
calculations of decay constants and form factors which are the
most relevant supporting measurements for $B$ physics. Second,
charm provides unique opportunities for indirect searches for
physics beyond the SM.

\section{Big Questions in Flavor Physics}

The big questions in quark flavor physics are: (1) ``What is the
dynamics of flavor?'' The gauge forces of the standard model (SM)
do not distinguish between fermions in different generations. The
electron, muon and tau all have the same electric charge, quarks
of different generations have the same color charge. Why
generations? Why three? (2) ``What is the origin of
baryogenesis?'' Sakharov gave three criteria, one is
$CP$-violation~\cite{Sakharov}. There are only three known
examples of $CP$-violation: the Universe, and the beauty and kaon
sectors. However, SM $CP$-violation is too small, by many orders
of magnitude, to give rise to the baryon asymmetry of the
Universe. Additional sources of $CP$-violation are needed. (3)
``What is the connection between flavor physics and electroweak
symmetry breaking?'' Extensions of the SM, for example
supersymmetry, contain flavor and $CP$-violating couplings that
should show up at some level in flavor physics but precision
measurements and precision theory are required to detect the new
physics.

\section{Charm in CKM physics}

  This is the decade of precision flavor physics. The goal is
to over-constrain the CKM matrix with a range of measurements in
the quark flavor changing sector of the SM at the per cent level.
If inconsistencies are found between, for example, measurements of
the sides and angles of the $B_d$ unitarity triangle, it will be
evidence for new physics. Many experiments will contribute
including BaBar and Belle, CDF, D0 at Fermilab, ATLAS, CMS, and
LHC-b at the LHC, BESIII, CLEO-c, and experiments studying rare
kaon decays.

\begin{figure}[t]
 \vspace{3.5cm}
\includegraphics{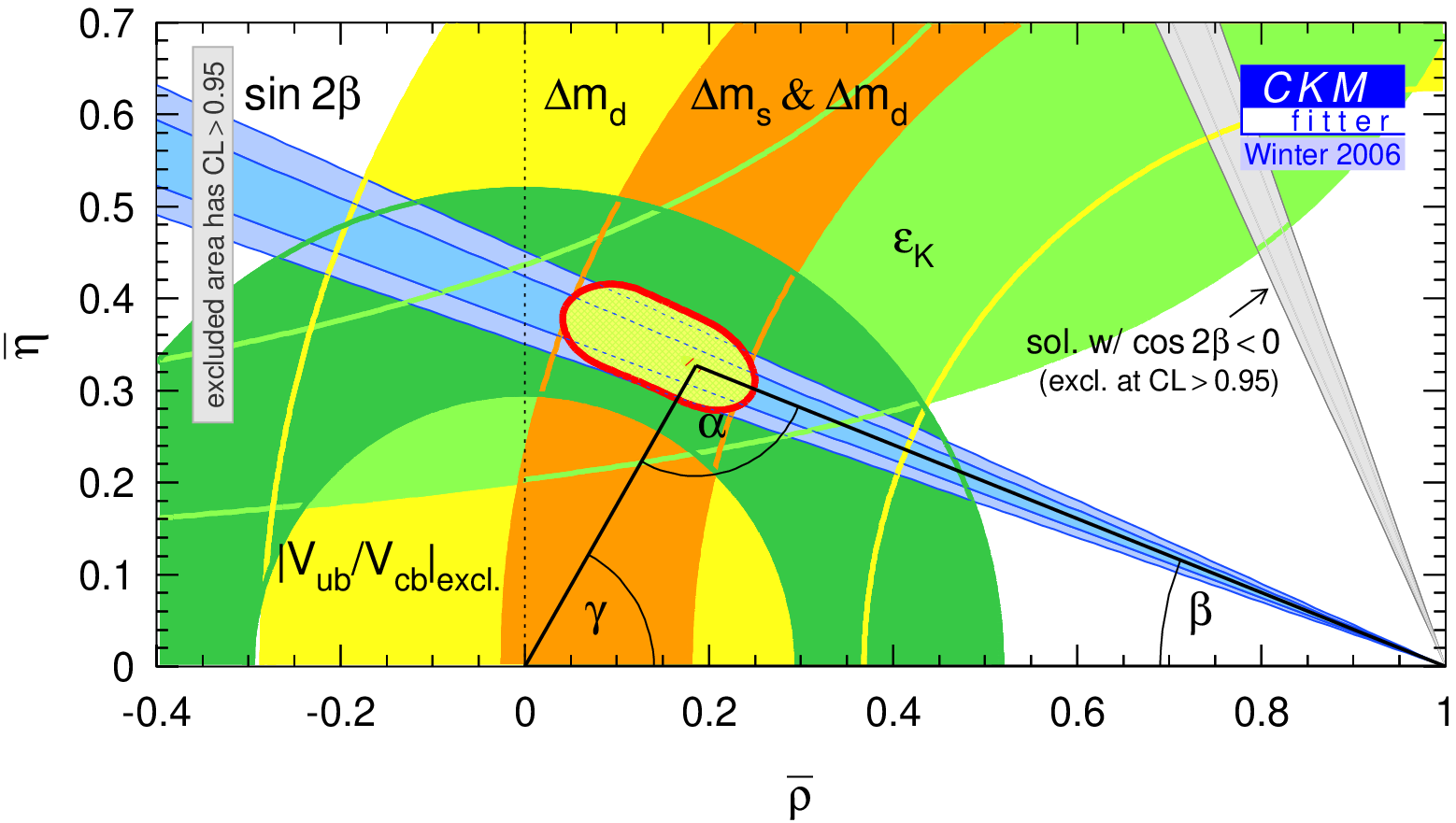}
\includegraphics{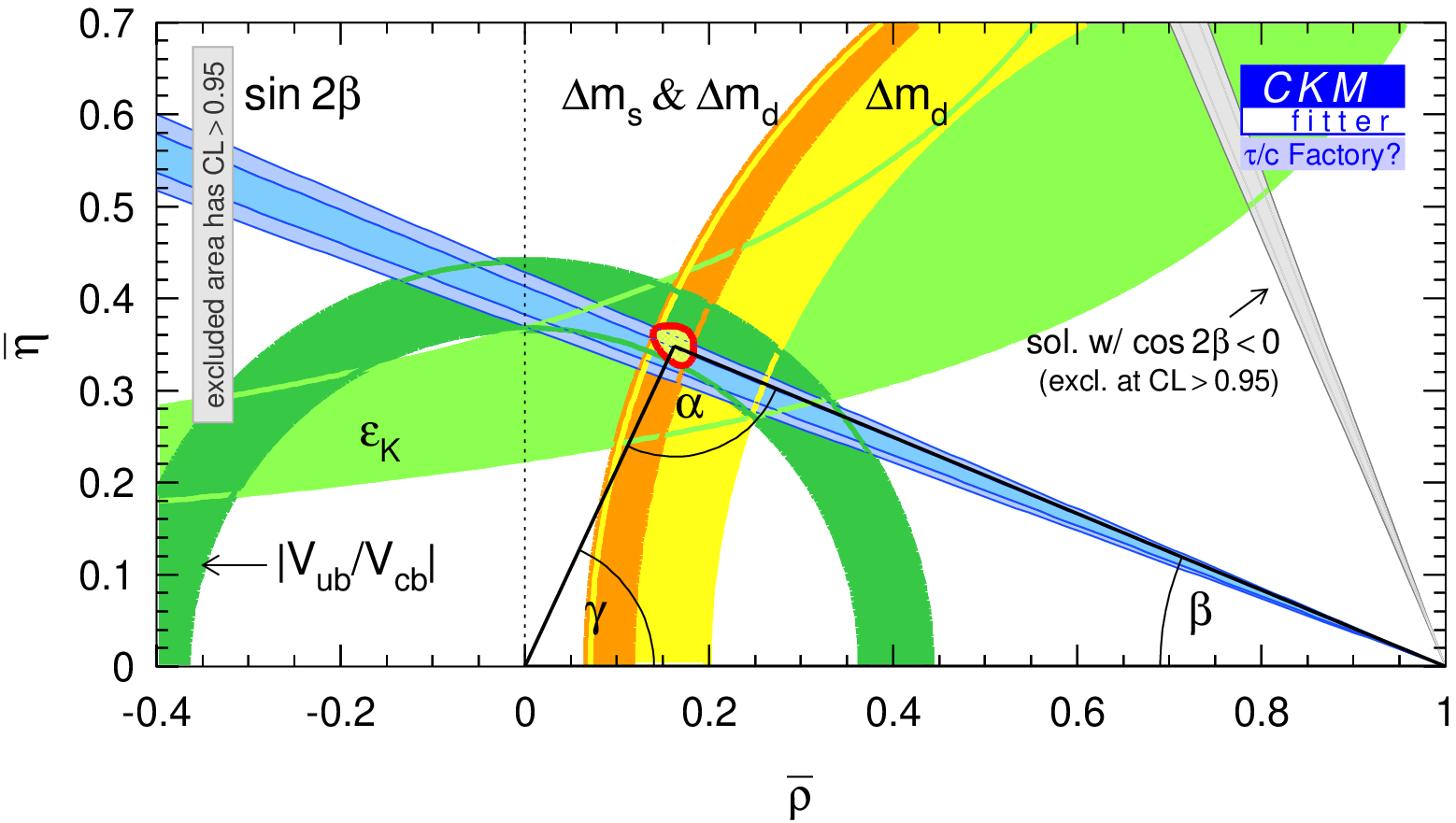}
 \caption{\it
      Lattice impact on the $B_d$ unitarity triangle from $B_d$
and $B_s$ mixing, $|\vub| / |\vcb|$, $\epsilon_K$, and $\sin 2
\beta$. (a) Winter 2006 status of the constraints including the
recent observation of $B_s$ mixing. (b) Prospects under the
assumption that LQCD calculations of $B$ system decay constants
and semileptonic form factors achieve the the precision projected
in Table~\ref{table:combine}.
    \label{CKM} }
\end{figure}

However, the study of weak interaction phenomena, and the
extraction of quark mixing matrix parameters remain limited by our
capacity to deal with non-perturbative strong interaction
dynamics. Current constraints on the CKM matrix are shown in
Fig.~\ref{CKM}(a). The widths of the constraints, except that of
$\sin 2 \beta$, are dominated by the error bars on the calculation
of hadronic matrix elements.
Recent advances in LQCD have produced calculations of
non-perturbative quantities such as $f_\pi$, $f_K$, and heavy
quarkonia mass splittings that agree with
experiment~\cite{Davies}.
Several per cent precision in charm and beauty decay constants and
form factors is hoped for, but the path to higher precision is
hampered by the absence of accurate charm data against which to
test lattice techniques. This is beginning to change with the BES
II run at the $\psi(3770)$, and the start of data taking at the
charm and QCD factory CESR-c/CLEO-c ~\cite{CLEO-cyellowbook}.
Later in the decade BES III at the new double ring accelerator
BEPC-II will also turn on~\cite{BESIII}. CLEO-c is in the process
of obtaining charm data samples one to two orders of magnitude
larger than any previous experiment, and the BES III data set is
expected to be $\sim \times 20$ larger than CLEO-c. These data
sets have the potential to provide unique and crucial tests of
LQCD, and other QCD technologies such as QCD sum rules and chiral
theory, with accuracies of 1-2\%.

If LQCD passes the charm factory tests,
we will have much greater confidence in lattice calculations of
decay constants and semileptonic form factors in $B$  physics.
When these calculations are combined with 500~fb$^{-1}$ of $B$
factory data, and improvement in the direct measurement of
$|\vtb|$ at the Tevatron~\cite{Swain}, they will allow a
significant reduction in the size of the errors on  $|\vub|,
|\vcb|, |\vtd| {\rm ~and~} |\vts|$, quantitatively  and
qualitatively transforming knowledge of the $B_d$ unitarity
triangle, see Fig.~\ref{CKM}(b), and thereby maximizing the
sensitivity of heavy quark physics to new physics.

Equally important, LQCD combined with charm data allows a
significant advance in understanding and control over
strongly-coupled, non-perturbative quantum field theories in
general. Field theory is generic, weak coupling is not. Two of the
three known interactions are strongly coupled: QCD and gravity
(string theory). Understanding strongly coupled theories may  be a
crucial to interpret new phenomena at the
high energy frontier.

\subsection{Decay Constants}

The $B_d$ $(B_s)$ meson mixing probability can be used to
determine $ |V_{td}|$ $(|V_{ts}|)$.
\begin{equation}
\Delta m_d \propto |V_{tb}V_{td}|^2 f_{B_d}^2 B_{B_d}
\end{equation}
The $B_d$ mixing rate is measured with exquisite precision
(1\%)~\cite{PDG2004} but the decay constant is calculated with a
precision of about 10-15\%. If theoretical precision could be
improved to 3\%, the error on $|V_{td}|$ would be about 5\%.

Since LQCD hopes to predict $f_B/f_{D^+}$ with a small error,
measuring $f_{D^+}$ would allow a precision prediction for $f_B$.
Hence a precision extraction of $|V_{td}|$ from the $B_d$ mixing
rate becomes possible. Similar considerations apply to $B_s$
mixing now it has been observed i.e. a precise determination of
$f_{D_s^+}$ would allow a precision prediction for $f_{B_s}$ and
consequently a precision measurement of $|\vts|$. Finally the
ratio of the two neutral $B$ meson mixing rates determines $|\vtd|
/ |\vts|$, but $|\vts| = |\vcb|$ by unitarity and $|\vcb|$ is
known to a few per cent, and so the ratio again determines $\vtd$.
Which method of determining $|\vtd|$ will have the greater utility
depends on which combination of hadronic matrix elements have the
smallest error.

Charm leptonic decays measure the charm decay constants
$f_{D_s^+}$ and  $f_{D^+}$ because $|V_{cs}|$ and $|V_{cd}|$ are
known from unitarity to 0.1\% and 1\% respectively.
\begin{equation}
{ {{\cal B}(D^+ \rightarrow \mu \nu_\mu) }\over {\tau_{D^+} } }=
{\rm (const.)} f_{D^+}^2 |V_{cd}|^2
\end{equation}
(Charge conjugation is implied throughout this paper.) The
measurements are also a precision test of the LQCD. At the start
of 2004 $f_{D^+}$ was experimentally undetermined and $f_{D_s^+}$
was known to 33\%.

\subsection{Semileptonic form factors}

$|V_{ub}|$ is determined from beauty semileptonic
decay
\begin{equation}
{{d\Gamma(B \rightarrow \pi e^- \bar\nu_e)} \over {dq^2}} = { \rm
(const.)} |V_{ub}|^2f_+^{B\pi}(q^2)^2
\end{equation}
The differential rate depends on a form factor, $f_+(q^2)$ that
parameterizes the strong interaction non-perturbative effects. A
representative value of $|\vub|$ determined from $B \rightarrow
\pi \ell^- \bar{\nu_e}$ is~\cite{HFAGSummer05}:
\begin{equation}
|V_{ub}| = (3.76 \pm 0.16^{+0.87}_{-0.51}) \times 10^{-3}
\end{equation}
where the uncertainties are experimental statistical and
systematic, and from the LQCD calculation of the form factor,
respectively.
The experimental errors are expected to be reduced to 5\% with $B$
factory data samples of $500 {\rm ~fb}^{-1}$ each, and the theory
error will dominate.

Again, because the charm CKM matrix elements are known from
unitarity, the differential charm semileptonic rate
\begin{equation}
{{d\Gamma(D \rightarrow \pi e^+ \nu_e)} \over {dq^2}} ={\rm
(const.)} |V_{cd}|^2f_+^{D\pi}(q^2)^2
\end{equation}
tests calculations of charm semileptonic form factors. Thus, a
precision measurement tests the LQCD calculation of the $D
\rightarrow \pi$ form factor. As the form factors governing $B
\rightarrow  \pi e^- \bar{\nu_e}$ and $D \rightarrow \pi e^+
\nu_e$ are related by heavy quark symmetry, the charm test gives
confidence in the accuracy of  the $B \rightarrow \pi$
calculation. The $B$ factories can then use a tested LQCD
prediction of the $B \rightarrow \pi$ form factor to extract a
precise value of $|V_{ub}|$.
At the start of 2004, ${\cal B} (D \rightarrow \pi e^+ \nu_e)$ had
been determined to 45\%~\cite{PDG2004,PDG2004_C}, and the absolute
value of the $D \rightarrow \pi$ form factor had not been
measured.

Lifetimes of the charm mesons are interpreted within the framework
of the Operator Product Expansion. Within OPE the total decay
width can be expressed as a series in $1/m_c$~\cite{Ikaros_OPE}.
\begin{eqnarray}
\Gamma(H_c) = \Gamma_{\rm c} + {\cal O} (1/m_c^2) \nonumber \\
+\Gamma_{\rm PI, WA, WS}(H_c) + {\cal O}(1/m_c^4)
\end{eqnarray}
Mechanisms in which light quarks in the $c-{\rm hadron}$ are
involved: Pauli Interference (PI), Weak Annihilation (WA) and Weak
Scattering (WS), are $O(1/m_c^3)$ ~but phase space enhanced. The
charm lifetimes are in Table~\ref{table:lifetimes}. The PDG2004
lifetimes are dominated by the exquisitely precise FOCUS
measurements from 2002. The $D^+$ and $D^0$ lifetimes are known to
$7$ and $4$ per mille, which is as precise as kaon lifetimes are
known.  PDG2004 does not include the $D_s$ lifetime measurement
from FOCUS and so we have averaged it with the PDG value in
Table~\ref{table:lifetimes}. The lifetimes can be explained within
OPE~\cite{Ikaros_OPE}. To gain a deeper understanding absolute
inclusive semileptonic branching ratios of $c-{\rm hadrons}$,
especially the $D_s^+$ and charm baryons, which are currently not
well known, need to be measured. For charm CKM physics, the most
important point to note is that errors on lifetimes are not a
limiting factor in the measurement of absolute rates.



\begin{table}[t]
\centering
\caption{ \it Charm lifetime world averages in fs.
}
\vskip 0.1 in
\begin{tabular}{@{}|ll|}
\hline
 Particle   & Lifetime (fs)  \\
\hline
$D^+$   &  $1040 \pm 7$  \\
$D_s^+$&  $504 \pm 4$  \\
$D^0$    & $410.3 \pm 1.5 $  \\
$\Xi_c^+$ & $442 \pm 26$ \\
$\Lambda_c^+$ & $200 \pm 6$ \\
$\Xi_c^+$ & $112^{+13}_{-10} $ \\
$\Omega_c^+$ & $ 69 \pm 12$ \\\hline
\end{tabular}
\label{table:lifetimes}
\end{table}

\section{Absolute Charm Branching Ratios}

We reviewed above the importance of  absolute charm leptonic and
semileptonic branching ratios. The absolute hadronic branching
ratios ${\cal B}(\dpkpp)$, ${\cal B}(D^0 \rightarrow K^-\pi^+)$,
and ${\cal B} (D_s^+ \rightarrow \phi \pi^+ )$ are also important
as, currently, all other $D^+$, $D^0$ and $D_s^+$ branching ratios
are determined from ratios to one or the other of these branching
fractions~\cite{PDG2004}. In consequence, nearly all branching
fractions in the $B$ and $D$ sectors depend on these reference
modes.

Absolute charm branching ratios are poorly known, see
Table~\ref{table:important_brs}. The reason is that charm produced
at B factories and at the Tevatron or at dedicated fixed target
facilities allows relative rate measurements but absolute rate
measurements are hard because backgrounds are sizeable, and,
crucially, the number of $D$ mesons produced is not easily
determined.

\begin{table}[tbp]
\centering
\caption{Status of important charm branching ratios circa 2004.}
\vskip 0.1 in
\label{table:important_brs}
\begin{tabular}{@{}|lll|}
\hline
 Mode  & ${\cal B}$ (\%) & $\delta{\cal B}/{\cal B}$  \\
\hline
$D^+ \rightarrow \mu^+ \nu_{\mu}$   &  $0.08^{+0.17}_{-0.05}$  & 100 \\
$D_s^+ \rightarrow \mu^+ \nu_{\mu}$&  $0.60 \pm 0.14$ & 24 \\
$D^0 \rightarrow \pi^- e^+ \nu_e$ & $ 0.30^{+0.23}_{-0.12}$ & 45 \\
$D^0 \rightarrow K^- \pi^+$ & $3.80 \pm 0.09$ & 2.4 \\
$D^+ \rightarrow K^- \pi^+ \pi^+$ & $9.2\pm 0.6$  &6.5  \\
$D_s^+ \rightarrow \phi \pi^+$ & $3.6 \pm 0.9 $ & 25 \\
$\Lambda_c^+ \rightarrow p K^- \pi^+ $ & $ 5.0 \pm 1.3$ & 26 \\
$J / \psi \rightarrow \mu^+ \mu^-$ & $5.88\pm 0.10$ & 1.7
\\\hline
\end{tabular}
\end{table}

To illustrate one way around this problem consider the clever
measurement of ${\cal B}(D_s^+ \rightarrow \phi \pi^+)$ from the
BABAR collaboration~\cite{BABAR-phipi}. The first stage in the
analysis is to produce a beam of $D_s^+$. This is achieved by
partially reconstructing $B^0 \rightarrow D_s^{*+} D^{*-}$, where
the $D^{*+}$ and the photon in the decay $D_s^{*+} \rightarrow
D_s^+ \gamma$ are reconstructed but the $D_s^+$ is not observed.
BABAR find
\begin{equation}
{\cal B}(B^0 \rightarrow D_S^{*+} D^{*-})=(1.88 \pm 0.09 \pm
0.17)\%
\end{equation}
In the second step $B^0 \rightarrow D_S^{*+} D^{*-}$ is fully
reconstructed
\begin{eqnarray}
{\cal B}(B^0 \rightarrow D_s^{*+} D^{*-}){\cal B} (D_s \rightarrow
\phi \pi)  \nonumber \\
 = (8.81 \pm 0.86) \times 10^{-4}
\end{eqnarray}
Dividing these
\begin{equation}
{\cal B} (D_s \rightarrow \phi \pi) = (4.81 \pm 0.52 \pm 0.38) \%
\end{equation}
The total error of 12.5\%, of which 7.5\% is systematic,
represents a dramatic improvement on  the 25\% precision of the
PDG value. Further improvement in the measurement of this
important quantity is expected at the B factories, although it
will be challenging to reduce the systematic error significantly.
In principle, a several per cent measurement of ${\cal B} (D_s
\rightarrow \phi \pi)$ is achievable at a charm factory.

\section{BES II and CLEO-c at the $\psi(3770)$}

In 2003 the venerable BES II detector accumulated an integrated
luminosity of  $33~{\rm pb}^{-1}$ at and around the $\psi(3770)$,
a factor three greater than the previous largest data sample
accumulated by Mark III in 1984. The Cornell Electron Storage Ring
(CESR) has been upgraded to CESR-c with the installation of 12
wiggler magnets to increase damping at low energies. The CLEO-c
detector is a minimal modification of the well understood CLEO III
detector. It is the first modern detector to operate at charm
threshold. In 2003 a CLEO-c pilot run accumulated $56~{\rm
pb}^{-1}$ at the $\psi(3770)$ ($360,000 D \bar D ~{\rm pairs}$)
and this was followed by the first full run accumulating a further
$225~{\rm pb}^{-1}$ for a total of $281~{\rm pb}^{-1}$at the
$\psi(3770)$ $(1.8 \times 10^6  D \bar D ~{\rm pairs})$ CLEO-c has
also accumulated about $200~{\rm pb}^{-1}$ at $\sqrt{s} \sim
4170$~MeV for $D_s$ physics. These $\psi(3770)$ datasets exceeds
those of the BESII (Mark III) experiments by factors of 30 (15).
CLEO-c expects to take data until April 2008 and will
approximately triple each data set by that time,

In the very near future the BEPCII Project will be commissioned,
This is a two ring machine with 93 bunches in each beam.
Luminosity is expected to be $10^{33} {\rm cm^{-2} s^{-1}}$ at
1.89 GeV $6 \times 10^{32} {\rm cm^{-2} s^{-1}}$ at 1.55 GeV and
$6 \times 10^{32} {\rm cm^{-2} s^{-1}}$ at 2.1 GeV. The linac was
installed in 2005. The ring is to be installed this year (2006)
and the BESIII detector will be in place and commissioned in 2007
with data taking beginning of 2008, with early running at the
$J/psi$. Although the detailed run plan has not been decided: an
example is given here. At 5/fb/yr or 15/fb/3yrs, there will be $90
\times 10^6  D \bar D ~{\rm pairs}$ or a factor 20 greater than
the full CLEO-c data sample. Three years at 4170 MeV would produce
$ 2 \times 10^6 D_s \bar{D_s} ~{\rm pairs}$ in three years again a
factor 20 greater than the full CLEO-c data set.

In the longer term proposed Super B Factories at KEK or SuperB or
a dedicated charm factory would produce an abundance of charm. For
example the SuperB machine at $10^{36} {\rm cm^{-2} s^{-1}}$ will
produce $10^{10} e^+ e^- \rightarrow  c \bar c {\rm ~pairs}/10^7
{\rm s}$. Due to the \lq \lq Linear Collider design" there is an
option to lower the energy to 4 GeV with a modest luminosity
penalty of a factor 10. In this mode of operation the super B
Factory becomes a super flavour factory. When discussing charm
factory results from CLEO-c I will extrapolate to BEPCII/BESIII
(my estimates, not official ones) and to super flavour. For the
latter I will assume $1 \times 10^{35} {\rm ~ for~} 10^7 {\rm ~s}$
which is  $(6.4 \times 10^9  D \bar D ~{\rm pairs})$ at the
$\psi(3770)$ exceeding the BEPCII and CESR-c data samples by a
factor of 70 and 1,000 respectively.

\subsection{Analysis Technique }

There are decisive advantages to running at charm threshold. As
$\psi \rightarrow D \bar D$, the technique is to fully reconstruct
one $D$ meson in a hadronic final state, the tag, and then to
analyze the decay of the second $D$ meson in the event to extract
inclusive or exclusive properties.

As $E_{\rm beam}=E_D$, the candidate is required to have energy
close to the beam energy, and the beam-constrained candidate mass,
$M_D\!\! =\!\! \sqrt{E_{{\rm beam}}^2 - p_{{\rm cand}}^2}$, is computed.
Charm mesons have many large branching ratios to low multiplicity
final states, and so the tagging efficiency is very high, about
25\%, compared to much less than 1\% for $B$ tagging at a $B$
factory.

\begin{figure}[btp]
 \vspace{5.0cm}
\includegraphics{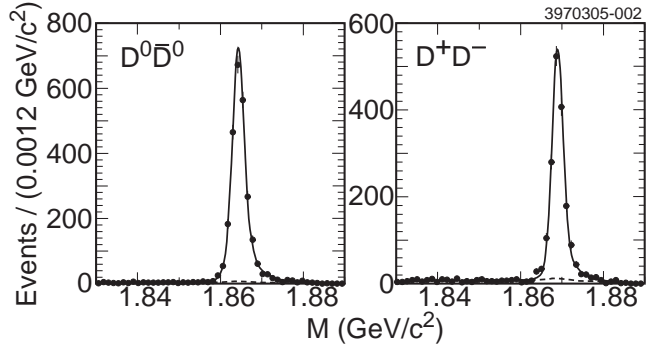}
 \caption{\it Beam constrained mass of $D$ mesons in
CLEO-c events in which both $D$ mesons have been fully
reconstructed.
    \label{Figure:double} }
\end{figure}

Tagging creates a single $D$ meson beam of known momentum.
The beam constrained mass for events in which the second $D$ meson
is also reconstructed are shown in Fig.~\ref{Figure:double}. These
double tag events, which are key to making absolute branching
fraction measurements, are pristine. The absolute branching
fraction is given by:
\begin{equation}
{\cal B} (D^+ \rightarrow K^- \pi^+ \pi^+) = { {N(K^- \pi^+
\pi^+)}\over { \epsilon (K^- \pi^+ \pi^+) \times N(D^-)}}
\end{equation}
where $N(K^- \pi^+ \pi^+)$ is the number of $D^+ \rightarrow K^-
\pi^+ \pi^+$ observed in tagged events, $\epsilon (K^- \pi^+
\pi^+)$ is the reconstruction efficiency and $N(D^-)$ is the
number of tagged events.

In a method similar to that pioneered by Mark
III~\cite{Balt,Adler}, CLEO fits to the observed single tag and
double tag yields for six $D^+$ and three $D^0$
modes~\cite{CLEO-had}. I will only consider the two most important
branching fractions here. For $D^0 \rightarrow K^- \pi^+$ the
total errors are comparable to previous measurements, see
Table~\ref{table:brDzero}. But the true improvement is that the
previous most precise measurements from ALEPH~\cite{ALEPH-KPI} and
CLEO~\cite{Akerib} were based on comparing $D^{*+} \to D^0 \pi_s,
D^0 \to K^- \pi^+$ with and without explicitly reconstructing the
$D^0$. The latter measurement relies on a correlation between the
momentum of the slow pion from the $D^*$ and the thrust axis of
the $e^+e^- \rightarrow q \bar q $ event. Consequently, these
early measurements had poor signal to noise whereas the CLEO-c
measurement has a signal to noise of about 60/1.

This is the most precise measurement of ${ \cal B} (\dpkpp) $ to
date, see Table~\ref{table:brDplus} but the improvement is again
much more than statistics. The previous most precise measurement,
which was from CLEO~\cite{Balest}, bootstrapped on $D^0
\rightarrow K^- \pi^+$ through a measurement of
\begin{equation}
{{ {\cal B }(D^{*+} \rightarrow D^0 \pi^+ ) {\cal B} (D^0
\rightarrow K^- \pi^+ )}  \over { {\cal B} (D^{*0} \rightarrow D^+
\pi^0){\cal B} (D^+ \rightarrow K^- \pi^+ \pi^+)}}
\end{equation}
so it was not independent of ${\cal B} (D^0 \to K^- \pi^+)$, while
the new measurement has no dependence on ${\cal B } (D^0 \to K^-
\pi^+)$ a much more satisfactory situation.

\begin{table}[tbp]
\centering
\caption{The $D^0 \rightarrow K^- \pi^+$ absolute charm branching
ratio. }
\vskip 0.1 in
 \label{table:brDzero}
\begin{center}
\begin{tabular}{@{}ll}
\hline
${\cal B} (\%)$                    & Error  (Source)  \\
\hline

$3.82\pm 0.07 \pm 0.12$    &  3.6\% (CLEO~\cite{Akerib})  \\
$3.90 \pm 0.09 \pm 0.12$   &  3.8\% (ALEPH~\cite{ALEPH-KPI}) \\
$3.80 \pm 0.09$                    & 2.4\% (PDG)    \\
$3.91 \pm0.08 \pm 0.09$                    &  3.1 \% (CLEO-c~\cite{CLEO-had}) \\
 \hline
\end{tabular}
\end{center}
\end{table}

\begin{table}[tbp]
\centering
\caption{The $D^+ \rightarrow K^- \pi^+ \pi^+$ absolute charm
branching ratio. }
\vskip 0.1 in
 \label{table:brDplus}
\begin{center}
\begin{tabular}{@{}ll}
\hline
${\cal B} (\%)$                    & Error  (Source)  \\
\hline

$9.3\pm 0.6 \pm 0.8$    &  10.8\% (CLEO~\cite{Balest})  \\
$9.1 \pm 1.3 \pm 0.4$   &  14.9\% (MKIII~\cite{MKIII-KPIPI}) \\
$9.1 \pm 0.7$                    & 7.7\% (PDG)    \\
$9.52 \pm0.25 \pm 0.27$                    &  3.9 \% (CLEO-c~\cite{CLEO-had}) \\
 \hline
\end{tabular}
\end{center}
\end{table}

BES II has performed a similar analysis.
These recent measurements are in remarkably good agreement with
the PDG averages, indicating that the charm, and hence beauty,
decay scales, are approximately correct and are now, finally, on a
solid foundation.


The CLEO-c $\psi(3770)$ integrated luminosity goal of $0.75~{\rm
fb^{-1}}$ may sound small compared to the more than $500~{\rm
fb^{-1}}$ collected by Belle, and the slightly smaller sample by
BABAR. However, the ability to perform a tagged analysis is
comparable at the two types of factory because the tagging
efficiency is at least 25 times larger at a charm factory than at
a $B$ factory, and the cross section is about six times larger.
Hence,
\begin{equation}
{ N(B~{\rm tags~at~a~}B~{\rm factory}) \over
  N(D~{\rm tags~at~a~charm~factory})   } \sim 1.
\end{equation}
In consequence the number of events in $100 {\rm pb^{-1}}$ with
two $D$ mesons reconstructed is about the same as the number of
events at 10 GeV with $500 {\rm fb^{-1}}$ with two $B$ mesons
reconstructed.
  Projections for the expected precision
with which the reference hadronic branching ratios will be
measured with a $0.75 {\rm fb^{-1}}$ data set are given in
Table~\ref{table:brproj}. CLEO-c and, later BES III, will set the
scale for all heavy quark measurements.

\begin{table}[tbp]
\centering
\caption{Charm factory hadronic branching ratio measurement
expected precision with $0.75 {\rm fb^{-1}}$ data samples at the
$\psi(3770)$ and above $D_s {\bar D_s}$ threshold. The first
uncertainty is statistical and the second systematic.}
\vskip 0.1 in
 \label{table:brproj}
\begin{center}
\begin{tabular}{@{}lll}
\hline
Mode                      & \multicolumn{2} {c} { $ \delta {\cal B} / {\cal B} $ (\%) } \\
& PDG2004 & $0.75 {\rm fb^{-1}}$  \\ \hline
$D^0 \rightarrow K^- \pi^+$    & 2.4\%  & 0.6\% , 1.1\%    \\
$\dpkpp $                        &  7.7\%  &  0.7\% ,  1.2\% \\
$D_s^+ \rightarrow \phi \pi $     & 12.5\%~\cite{BABAR-phipi}  & 4.0\%    \\
 \hline
\end{tabular}
\end{center}
\end{table}

\subsection{Charm Decay Constant }

\begin{figure}[btp]
 \vspace{6.0cm}
\includegraphics{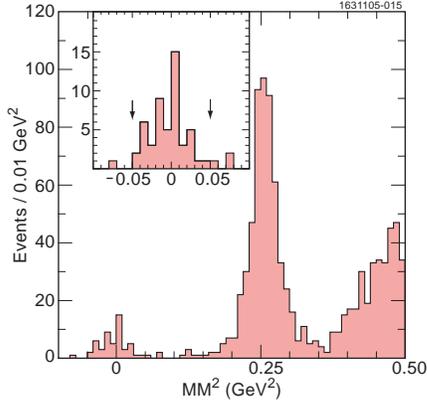}
 \caption{\it The $MM^2$ distribution in events with
$D^-$ tag, a single charged track of the correct sign, and no
additional (energetic) showers. The insert shows the signal region
for $D^+ \rightarrow \mu \nu_\mu$. A $\pm 2 \sigma$ range is
indicated by the arrows.  }
\label{missmass}
\end{figure}

The measurement of the leptonic decay $D^+ \rightarrow \mu^+
\nu_\mu$ benefits from the fully tagged $D^-$ at the $\psi(3770)$.
One observes a single charged track recoiling against the tag that
is consistent with a muon of the correct sign. Energetic
electromagnetic showers un-associated with the tag are not
allowed. The missing mass $MM^2 = m_{\nu}^2$ is computed; it 
peaks at zero for a decay where only a neutrino is unobserved.
Fig.~\ref{missmass} shows the $MM^2$ distribution from
CLEO-c~\cite{CLEO_leptonic}.

There are 50 candidate signal events, and $2.81 \pm
0.3^{+0.84}_{-0.22}$ background events. After correcting for
efficiency, CLEO-c finds
\begin{equation}
{\cal B} (D^+ \rightarrow \mu^+ \nu_\mu) = (4.40 \pm
0.66^{+0.09}_{-0.12}) \times 10^{-4},
\end{equation}
where the uncertainties are statistical and systematic,
respectively. Under the assumption of three generation unitarity,
and using the precisely known $D^+$ lifetime, CLEO-c obtains
\begin{equation}
f_{D^+} = (222.6 \pm 16.7^{+2.8}_{-3.4} ) {\rm ~MeV}.
\end{equation}
This is the most precise measurement of
$f_{D^+}$~\cite{CLEO_leptonic}. The result appeared at
Lepton-Photon 2005 just two days after the first unquenched
lattice QCD calculation~\cite{Aubin-decay} had predicted:
\begin{equation}
f_{D^+} = (201 \pm 3 \pm 17 ) {\rm ~MeV}.
\end{equation}
The combined experimental error is 8\% while the LQCD error is
also 8\%~\cite{Aubin-decay}. The results are in good agreement but
errors are still large. The only other positive observation of
this decay is by BES II who found three candidate events with a
background of 0.25 events in their $33{\rm pb}^{-1}$ data sample.
They find a branching ratio of
$(0.122^{+0.111}_{-0.053}\pm0.010)\%$ corresponding to $f_{D^+}
=(371^{+129}_{-119}\pm 25) {\rm ~MeV}$~\cite{BESII_f}. The CLEO
value is considerably smaller and in better agreement with
expectations from the lattice and other theoretical approaches.
With $0.75{\rm
~fb^{-1}}$ a 4.5\% error for $f_{D^+}$ is expected. Similar
precision is expected for $f_{D_s^+}$ at $\sqrt{s}= 4160$~MeV. BES
III will make even more precise measurements achieving a precision
of several per cent for both $f_{D^+}$ and $f_{D_s^+}$ which is
well matched to the ultimate precision of the LQCD calculations.

\begin{figure}[btp]
 \vspace{6.0cm}
\includegraphics{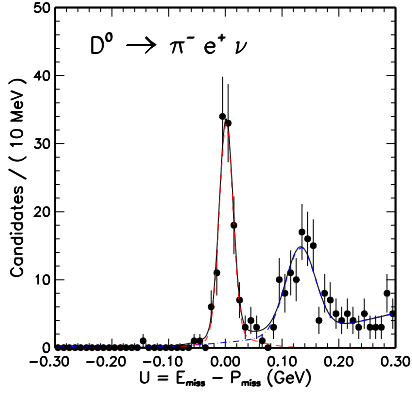}
 \caption{\it The $U=E_{miss}-P_{miss}$ distribution in
events with a $\bar{D^0}$ tag, a positron, and a single charged
track of the correct sign. The peaks at zero and 0.13~GeV
correspond to $D^0 \rightarrow \pi^- e^+ \nu_e$ and $D^0
\rightarrow K^- e^+ \nu_e$ (preliminary.) }
\label{fig:CLEO_pienu}
\end{figure}

\subsection{Measurement of the Charm Semileptonic Form Factors }

The measurement of semileptonic decay absolute branching ratios
and absolute form factors is also based on the use of tagged
events.
%
The analysis procedure, using $D^0 \rightarrow \pi^- e^+ \nu_e$ as
an example is as follows. A positron and a hadronic track are
identified recoiling against the tag. The quantity $U = E_{miss}-
P_{miss}$ is calculated, where $E_{miss}$ and $P_{miss}$ are the
missing energy and missing momentum in the event. For a tagged
event with a semileptonic decay $E_{miss}$ and $P_{miss}$ are the
components of the four-momentum of the neutrino. $U$ peaks at zero
if only a neutrino is missing. The $U$ distribution in $56 {\rm
~pb^-1}$ of CLEO-c data is shown in Fig.~\ref{fig:CLEO_pienu}
where a clean signal of about 100 events is observed for $D
\rightarrow \pi e^+ \nu_e$ with
$S/N~20/1$~\cite{CLEO_semileptonic}. In previous analyses at $B$
Factories and fixed target experiments the background was usally
larger than the signal see for example~\cite{Hsu}.

The kinematic power of running at threshold also allows previously
unobserved modes such as $D^0 \rightarrow \rho^- e^+ \nu_e$ to be
easily identified~\cite{CLEO_semileptonic}.
BES II have performed similar
analyses~\cite{BES_more_semileptonic}~\cite{BES_semileptonic} and
results are in good agreement with CLEO-c. Selected CLEO-c
absolute semileptonic branching ratio measurements are compared to
PDG values in Table~\ref{table:slbr}.
\begin{table}[tbp]
\centering
\caption{Selected CLEO-c charm semileptonic branching ratio
measurements in \% and a comparison to the PDG.}
\vskip 0.1 in
\label{table:slbr}
\begin{center}
\begin{tabular}{@{}lll}
\hline
Mode                      & PDG & CLEO-c \\
\hline
$\dpie $          & $0.36 \pm 0.06$           &  $0.26 \pm 0.03 \pm 0.1 $ \\
$\dke $           &$3.58 \pm 0.18 $         & $3.44 \pm 0.10 \pm 0.1$    \\
$D^+ \rightarrow \pi^0 e^+ \nu_e $  & $0.31 \pm 0.05$ & $0.44 \pm 0.06 \pm 0.01$   \\
$ D^+ \rightarrow \overline K^0 e^+ \nu_e$ &  $6.7 \pm 0.9$ & $8.71 \pm 0.38 \pm 0.37$  \\
\hline
\end{tabular}
\end{center}
\end{table}

This modest data sample has already produced several important
measurements. The ratio of $\Gamma(D^0 \rightarrow
K^-e^+\nu_e)/\Gamma(D^+ \rightarrow \overline K^0 e^+ \nu_e)$ is
expected to be unity by isospin. The PDG value is $1.35 \pm
0.19$~\cite{PDG2004}. Using the measured branching fractions for
the decays of $D^0 \rightarrow K^-e^+\nu_e$ and $D^+ \rightarrow
\overline K^0 e^+ \nu_e$ and the lifetimes of the $D^0$ and
$D^+$~\cite{PDG2004}  CLEO-c obtains the ratio of the decay widths
\begin{equation}
\frac{\Gamma(D^0 \rightarrow K^-e^+\nu_e)} {\Gamma(D^+ \rightarrow
\overline K^0 e^+\nu_e)} = 1.00 \pm 0.05 \pm 0.04
\end{equation} where the first
error is statistical and the second systematic. The CLEO0-c
result, and a less precise result from BES II,  are consistent
with unity thereby solving a long standing puzzle.

\begin{figure}[btp]
 \vspace{5.8cm}
\includegraphics{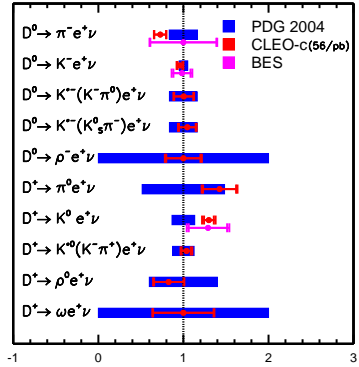}
 \caption{\it CLEO-c, BES II and the PDG values for a
range of charm meson semileptonic branching ratios. Results are
normalized to PDG values for ease of display.}
\label{fig:CLEO-SL}
\end{figure}

As the charm CKM matrix elements are known from unitarity, the
absolute differential charm semileptonic rate
\begin{equation}
{{d\Gamma(D \rightarrow \pi e^+ \nu_e)} \over {dq^2}} ={\rm
(const.)} |V_{cd}|^2f_+(q^2)^2
\end{equation}
tests calculations of charm semileptonic form factor $q^2$
dependence and form factor magnitude.  A precision absolute
branching fraction measurement also tests the magnitude of the
form factor if an assumption is made about the functional form of
the $q^2$ dependence. Recently there have been several beautiful
measurements of the form factor shape in $D \rightarrow K \ell^+
\nu_{\ell}$ and  $D \rightarrow \pi \ell^+ \nu_{\ell}$ by  CLEO,
FOCUS, Belle, and BABAR.  By reconstructing two $D$ mesons in
$e^+e^- \rightarrow  c \bar c $ events at 10 GeV  Belle are able
to make an absolute measurement and so a determination  of the
form factor magnitude as well. CLEO-c promise results soon.

In a pseudoscalar to pseudoscalar transition the differential rate
is proportional to the third power of the daughter hadron momentum
due to the P-wave nature of the decay. The $p^3$ term dominates
the differential rate. The form factor parameterizes the
additional $q^2$ dependence of the semileptonic amplitude arising
from non-perturbative QCD. The form factor is largest at $q^2 =
q^2_{max}$ where the daughter hadron is stationary in the rest
frame of the $D$ meson and decreases by about a factor of two at
$q^2=0$. Since most of the rate is at $q^2=0$ it is traditional to
normalize the form factor at $q^2=0$, however it is simpler to
calculate the form factor at $q^2=q^2_{\max}$ where the rate
vanishes as we at the edge of phase space.

\begin{figure}[btp]
 \vspace{5.0cm}
\includegraphics{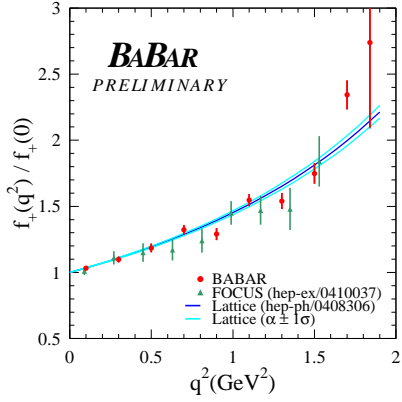}
\caption{\it The differential rate, normalized to the
rate at $q^2=0$ for the decay $D^0 \rightarrow K^- \ell
\nu_{\ell}$ after removal of phase space factors, compared to the
LQCD prediction. } \vskip -1.5em

\label{FOCUS-overlay}
\end{figure}

Several choices for the functional form  of $ f_+(q^2)$   have
been proposed. The simple pole model is a form predicted by vector
meson dominance also called nearest pole dominance~\cite{VMD}, in
which exchange is dominated by the lowest lying vector meson (the
spectroscopic pole) with the quantum numbers of the $c \rightarrow
s $ transition.
\begin{equation}
f_+(q^2) =   {  {f_+(q^2=0)} \over { 1 - { q^2 \over m^2_{pole}}}
}
\end{equation}
Where $m_{pole}=M_{D^*}$ for $D \rightarrow \pi  \ell \nu_{\ell}$
and $m_{pole}=M_{D^*_s}$ for $D \rightarrow K  \ell \nu_{\ell}$.
At lower values of $q^2$ the spectrum has contributions from
higher poles, and to account for this the modified pole or BK
parametrization was proposed~\cite{BK}
\begin{equation}
f_+(q^2) =   {  {f_+(q^2=0)} \over { (1 - { q^2 \over m^2_{pole}})
(1 - { \alpha q^2 \over m^2_{pole}}) } }
\end{equation}
Here $\alpha$ parameterizes the contributions of all additional
poles combined, and $m_{pole}$ remains the spectroscopic pole. The
$q^2$ spectrum in $D \to K \ell \nu_{\ell}$ can be described by
the pole model within experimental resolution, but the pole mass
needed to do so is far from the spectroscopic pole. The B-K
parametrization describes the data for $D \rightarrow K \ell
\nu_{\ell}$ within the experimental precision and also provides a
way to parameterize the lattice calculations. A comparison of  a
lattice prediction for $\alpha$ to data is shown for FOCUS and
BABAR data in Figure~\ref{FOCUS-overlay}. The precision of the
prediction and the measurements are at the 10\% level. Agreement
is good, although the errors are still large.

\begin{table}[tbp]
\centering
\caption{Experimental measurements and theoretical predictions  of
shape parameters in $D \rightarrow K$ semileptonic decay.}
\vskip 0.1 in
\label{table:alpha-DK}
\begin{center}
\begin{tabular}{@{}lll}
\hline
Measurement                      & $\alpha$ & $m_{pole}$ GeV \\
\hline
E691 1989~\cite{E691-89}         & --          &  $2.1^{+0.4}_{-0.2}\pm 0.2 $ \\
CLEO 1991~\cite{CLEO-91}         & --         & $2.0^{+0.4+0.3}_{-0.2-0.2}$    \\
MARKIII 1991~\cite{MARKIII-91}   & --     & $1.8^{+0.5+0.3}_{-0.2-0.2}$   \\
CLEOII 1993~\cite{CLEOII-93}   &  -- & $2.00 \pm 0.12 \pm 0.18$  \\
E687 1995~\cite{E687-95} &--& $1.87^{+0.11+0.07}_{-0.08-0.06}$ \\
CLEOIII 2005~\cite{Hsu} & $0.36\pm0.10^{+0.03}_{-0.07}$ & $1.89 \pm 0.05^{+0.04}_{-0.03}$\\
FOCUS 2005~\cite{FOCUS-05} & $0.28 \pm 0.08 \pm 0.07$ & $1.93 \pm
0.05 \pm0.03$ \\
Belle 2006~\cite{Belle-06} & $0.52 \pm 0.08 \pm 0.06$ & -- \\
BABAR 2006~\cite{BABAR-06} &$0.43 \pm 0.03 \pm 0.04 $ &  $1.854
\pm 0.016 \pm 0.020$ \\
LQCD~\cite{Aubin} & $0.50 \pm  0.06 \pm 0.07$ & -- \\
LCSR~\cite{LCSR-alpha} & $-0.07^{+0.15}_{-0.07}$ & --\\
CQM~\cite{CQM-alpha} & $0.24$ & -- \\
\hline
\end{tabular}
\end{center}
\end{table}

\begin{figure}[btp]
 \vspace{6.0cm}
\includegraphics{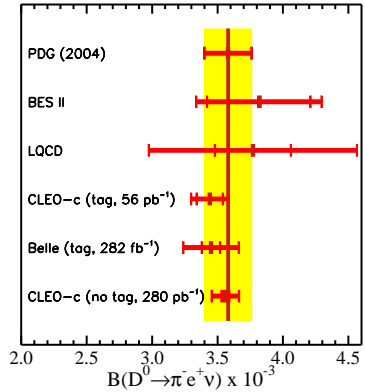}
\caption{\it Measurements of the absolute branching
fraction for $D \rightarrow \pi e^+ \nu_e$ and comparison to LQCD.
A preliminary result from an untagged measurement from CLEO-c has
also been included. }
\label{Norm-display}
\end{figure}

The FOCUS, BABAR and Belle measurements check the shape of the
form factor. The normalization can be checked by either fitting to
the differential rate to obtain $f^+(0)V_{cx}$ or from the
absolute branching fraction, in both cases using unitarity and the
$D$ meson lifetime. A comparison of absolute branching fraction
measurements and the LQCD prediction is shown in
Figure~\ref{Norm-display}. Here while the measurement has recently
become much more precise, the precision of the prediction lags
experiment significantly. Agreement is reasonable, although the
theory errors are in urgent need of being reduced.

The $q^2$ resolution at a charm factory is about 0.025 GeV$^{2}$,
which is more than a factor of 10 better than CLEO III which
achieved a resolution of 0.4 GeV$^{2}$~\cite{Hsu}. This huge
improvement is due to the kinematics at the $\psi(3770)$
resonance, i.e. that the $D$ meson momentum is known. (Belle have
recently achieved similar $q^2$ resolution by using a charm
tagging technique at 10 GeV.) The combination of large statistics,
and excellent kinematics will enable the absolute magnitudes and
shapes of the form factors in every charm semileptonic decay to be
measured, in many cases to a precision of a few per cent. This is
a stringent test of LQCD.

By taking ratios of semileptonic and leptonic rates, CKM factors
can be eliminated. Two such ratios are
\begin{eqnarray}
{\Gamma(D^+ \rightarrow \pi^0 e^+ \nu_e)} / {\Gamma (D^+
\rightarrow \mu \nu_\mu)} \nonumber \\
{\Gamma(D_s^+ \rightarrow (\eta {\rm ~or~} \phi ) e^+ \nu_e)} /
{\Gamma (D_s^+ \rightarrow \mu \nu_\mu)}
\end{eqnarray}
These ratios depend purely on hadronic matrix elements and can be
determined to 8\% and so will test amplitudes at the 4\% level.
This is an exceptionally stringent test of LQCD.

If LQCD passes the experimental tests outlined above it will be
possible to use the LQCD calculation of the $B \rightarrow \pi$
form factor with increased confidence at the $B$ factories to
extract a precision $\vub$ from $B \rightarrow  \pi e^-
\bar\nu_e$.  BaBar and Belle will also be able to compare the LQCD
prediction of the shape of the $B \rightarrow \pi$ form factor to
data as an additional cross check.

\begin{table}[tbp]
\centering
\caption{Experimental measurements and theoretical predictions  of
shape parameters in $D \rightarrow \pi$ semileptonic decay.}
\vskip 0.1 in
\label{table:alpha-Dpi}
\begin{center}
\begin{tabular}{@{}lll}
\hline
Measurement                      & $\alpha$ & $m_{pole}$ GeV \\
\hline
CLEOIII 2005~\cite{Hsu} &$0.37^{+0.20}_{-0.31}\pm{0.15}$ & $1.86^{+0.10+0.10}_{-0.09-0.03}$ \\
FOCUS 2005~\cite{FOCUS-05} & --  & $1.91^{+0.03}_{-0.15}\pm 0.07$ \\
Belle 2006~\cite{Belle-06} & $0.10 \pm 0.21 \pm 0.10$ & -- \\
LQCD~\cite{Aubin} & $0.44 \pm  0.06 \pm 0.07$ & -- \\
LCSR~\cite{LCSR-alpha} & $0.01^{+0.11}_{-0.07}$ & --\\
CQM~\cite{CQM-alpha} & $0.30$ & -- \\
\hline
\end{tabular}
\end{center}
\end{table}

Successfully passing the experimental tests allows the charm
factories to use LQCD calculations of the charm semileptonic form
factors to directly measure $|\vcd|$ and $|\vcs|$. Using the
isospin averaged semileptonic widths $\Gamma(D \rightarrow K e^+
\nu_e)$ and $\Gamma (D \rightarrow \pi e^+ \nu_e)$
~\cite{CLEO_semileptonic} and the LQCD prediction of the
semileptonic partial width~\cite{Aubin} I obtain
\begin{eqnarray}
V_{cs} = 0.957 \pm 0.017 \pm 0.093
\nonumber \\
 V_{cd} = 0.213
\pm 0.008 \pm 0.029
\end{eqnarray}
where the uncertainties are experimental statistical, experimental
systematic and from LQCD.  The results are consistent with the
unitarity values
\begin{eqnarray}
V_{cs} = 0.9745 \pm 0.0008
\nonumber \\
 V_{cd} = 0.2238 \pm 0.012
\end{eqnarray}
$V_{cd}$ has previously been determined from neutrino production
of di-muons off of nucleons, and $V_{cs}$ has been determined from
$W \rightarrow cs$ transitions at LEP to be~\cite{PDG2004}
\begin{eqnarray}
|V_{cs}| = 0.976 \pm 0.014
\nonumber \\
 |V_{cd}| = 0.224 \pm 0.012
\end{eqnarray}
Due to the large theoretical uncertainties in the CLEO-c numbers
the extracted values of $V_{cs}$ and $V_{cd}$ should be considered
as tests of LQCD. Nonetheless, they are the single most precise
determinations of $V_{cs}$ and $V_{cd}$ to date. With $0.75 {\rm
fb^{-1}}$ of data the CLEO-c precision is expected to be
respectively:
\begin{eqnarray}
|V_{cs}| = \sqrt{0.8\% \oplus \delta \Gamma / 2\Gamma}
\nonumber \\
| V_{cd}| = \sqrt {1.6\% \oplus  \delta \Gamma / 2\Gamma}
\end{eqnarray}
Where $\delta \Gamma/\Gamma$ is the
uncertainty in the partial rate from theory.  This in turn allows
new unitarity tests of the CKM matrix. For example, the second row
of the CKM matrix can be tested at the few \% level. With the
current measurements I find:
\begin{equation}
1-(|V_{cs}|^2 + |V_{cd}|^2 + |V_{cb}|^2 )= 0.037 \pm 0.181
\end{equation}
which is consistent with unitarity, with an uncertainty dominated
by the LQCD charm semileptonic form factor magnitude The
measurements also allow the first column of the CKM matrix to be
tested with similar precision to the first row (which is currently
the most stringent test of CKM unitarity); finally, the ratio of
the long sides of the $uc$ unitarity triangle will be tested to a
few percent.

Table~\ref{table:combine} provides a summary of projections for
the precision with which  the CKM matrix elements will be
determined if LQCD passes the charm factory tests in the $D$
system. In the tabulation the current precision of the CKM matrix
elements is obtained by considering methods applicable to LQCD,
for example the determination of $|\vcb|$ and $|\vub|$ from
inclusive decays and OPE is not included. The projections are made
assuming $B$ factory data samples of 500~fb$^{-1}$ and improvement
in the direct measurement of $|\vtb|$ expected from the Tevatron
experiments~\cite{Swain}.

\begin{table}[tbp]
\centering
\caption{LQCD impact (in per cent) on the precision of CKM matrix
elements. A charm factory data set of 3/fb and a B factory data
set of 500/fb is assumed.}
\vskip 0.1 in
 \label{table:combine}
\centering
\begin{tabular}{lrrrrrr}
\hline
  ~~ & $\vcd$ & $\vcs$ & $\vcb$ & $\vub$ & $\vtd$ & $\vts$ \\
\hline
2004  & 7 & 11 & 4 & 15 & 36 & 39   \\
LQCD & 2 & 2 & 3 & 5 & 5 & 5 \\
 \hline
\end{tabular}
\end{table}

\subsection{The bottom line}

How can we be sure that if LQCD works for $D$ mesons it will work
for $B$ mesons? Or, equivalently, is charm factory data sufficient
to demonstrate that lattice systematic errors are under control?
There are a number of reasons to answer this question in the
affirmative. (1) There are two independent effective field
theories: NRQCD and the Fermilab method. (2) The CLEO-c, and later
BESIII,  data provide many independent tests in the $D$ system;
leptonic decay rates, and semileptonic modes with rate and shape
information. (3) The $B$ factory data provide additional
independent cross checks such as $ { d \Gamma(B \rightarrow \pi
\ell \nu ) / d p_\pi}$. (4) Unlike models, methods used for the
$D/B$ system can be tested in heavy onia with measurements of
masses, and mass splittings, $\Gamma_{ee}$ and electromagnetic
transitions. (5) The main systematic errors limiting accuracy in
the $D/B$ systems are: chiral extrapolations in $m {\rm light}$,
perturbation theory, and finite lattice spacing. These are similar
for charm and beauty quarks. In my opinion a combination of CLEO-c
and BES III data in the $D$ systems and onia, plus information on
the light quark hadron spectrum, can clearly establish whether or
not lattice systematic errors are under control.

While this picture is encouraging, experimentalists also have
concerns. The lattice technique is all encompassing but LQCD
practitioners are very conservative about what can be calculated.
For example when there was a hint that  $\sin 2\beta (\psi K_S^0)
\neq \sin 2 \beta (\phi K_S^0 )$,  and when CP violation was
observed in $B \rightarrow K \pi$~\cite{ACP}  the lattice was not
able to contribute. There is a pressing need to move beyond the
limited set of easy to calculate quantities in the next few years:
for example resonances such as $\rho, ~\phi {\rm ~and~} K^*$ may
be difficult to treat on the lattice, but they feature in many
important $D$ semileptonic decays which will be well measured by
the charm factories.  There is also a need to be able to calculate
for states near threshold such as $\psi(2S)$ and $D_s(0)^+$, and
hadronic weak decays in the $B$ and $D$ systems as well.

\section{New physics searches with charm}

In the early part of the 20th Century table top nuclear $\beta$
decay experiments conducted at the MeV mass scale probed the $W$
at the 100 GeV mass scale. In an analogous way can we find
violations of the Standard Model by studying low energy processes?
The existence of multiple fermion generations appears to originate
at very high mass scales and so can only be studied indirectly.
Mixing, CP violation, and rare decays may investigate the new
physics at these scales through intermediate particles entering
loops. Why is charm a good place to look? In the charm sector, the
SM contributions to these effects are small, in other words, a
background free search for new physics is possible (see caveats
below). Typically $D^0 - \bar{D^0}$ mixing ${\cal O} (<10^{-2})$,
CP asymmetry ${\cal O} (<10^{-3})$ and rare decays ${\cal O}
(<10^{-6})$. In addition, charm is a unique probe of the up-type
quark sector (down quarks in the loop). The sensitivity of
searches for new physics in charm depends on high statistics
rather than high energy.

\subsection{Charm Mixing}

Mixing
has been a fertile ground for discoveries. The neutral kaon mixing
amplitude occurs at the same order as the kaon decay width
$\propto |\vus|^2$ and so the mixing rate is of order unity. The
mixing rate, which vanishes in the SU(4) symmetry limit, was
measured in 1958, was used to bound the charm quark mass, 16 years
before the discovery of charm. The CP violating part of $K^0
\bar{K^0}$ mixing, $\epsilon_K$, first measured in 1964 was a
crucial clue that the top quark existed, thirty years before its
discovery. In the $B^0 \bar{B^0}$ system the top quark dominates
the mixing amplitude, the $B$ decay width is Cabibbo suppressed
$\propto |\vcb|^2$ and mixing is also Cabibbo suppressed $\propto
| \vtd|^2$. The mixing rate is again of order unity, which was an
early indication that $m_{\rm top}$ was large. In $D^0\bar{D^0}$
mixing the amplitude is proportional to $\sin^2 \theta_c \sim
0.05$ but the decay width is not Cabibbo suppressed $( \vcs \sim
1)$. There is additional GIM suppression of order $ ( m_s^2 -
m_d^2)/m_W^2 = 0$ in the SU(3) limit, and so the rate for $D$
mixing in the SM is the product of Cabibbo suppression and an
SU(3) breaking term, the latter being extremely difficult to
estimate~\cite{mix-many}
\begin{equation}
{\rm mixing} \sim \sin^2\theta_c \times [SU(3) {\rm~breaking}]^2
\end{equation}


\begin{table}[tbp]
\centering
\caption{Summary of measurements of $y_{CP}$}
\vskip 0.1 in
\begin{center}

\label{table:mix-y}
\begin{tabular}{@{}|lll|}
\hline
                      & \multicolumn{2} {c|} { \% } \\
\hline
Belle & 2003 & $1.15 \pm 0.69 \pm 0.38$ \\
BABAR &2003  & $0.8 \pm 0.4^{+0.5}_{-0.4}$   \\
CLEO  & 2001  &  $-1.1 \pm 2.5 \pm 1.4$  \\
Belle & 2001 & $-0.5 \pm 1.0^{+0.7}_{-0.8}$ \\
FOCUS & 2000 & $3.4 \pm 1.4\pm 0.7$ \\
E791 & 1996 & $0.8 \pm 2.9 \pm 1.0$  \\
 \hline
\end{tabular}
\end{center}
\end{table}

In consequence, SM predictions span the range bounded by the
experimental upper limit of 1\% and the short distance box diagram
rate of $ {\cal O}(10^{-8})$~\cite{Datta} and the di-penguin rate
$ {\cal O}(10^{-10})$~\cite{dipenguin}. New physics predictions
span the same large range~\cite{A-Petrov}, implying that the
observation of $D$ mixing alone is not a clear indication of new
physics. However, the current experimental bounds $ {\cal
O}(10^{-2})$~\cite{Bianco,Annrev,Asner} already constrain new
physics models.

Neutral meson mixing is characterized by two dimensionless
parameters
\begin{equation}
x = \Delta M / \Gamma , y = \Delta \Gamma /  2 \Gamma
\end{equation}
where $\Delta m = m_1-m_2$ is the mass difference and $\Delta
\Gamma = \Gamma_1-\Gamma_2$ is the width difference between the
two neutral $D$ meson $CP$ eigenstates, and $\Gamma$ is the
average width. If mixing occurs either $x$ or $y$ or both are
non-zero.

The lifetime difference $y$ is constructed from the decays of a
$D$ into physical states, and so it is expected to be dominated by
SM contributions. In addition to the tiny SM contribution, the
mass difference, $x$, is sensitive to new particles in the box
diagram loop. Thus, new physics can significantly modify $x$,
leading to $x
>>y$. This signature for new physics is lost, however, if a
relatively large $y$ of ${\cal O}(1\%)$ is observed~\cite{Nir}. As
CP violating effects in mixing in the SM must involve the third
quark generation, and since the bottom quark contribution to the
box diagram is highly suppressed, $\propto \vcb \vub^*$, the
observation of $CP$ violating effects in $D$ mixing would be an
unambiguous signal of new physics.

\begin{table}[tbp]
\centering
\caption{Summary of searches for $D^0$ mixing with semileptonic
decays. (Limits are 90\% C.L.)}
\vskip 0.1 in
\begin{center}
 \label{table:mix-SL}
\begin{tabular}{@{}|lll|}
\hline
                      & \multicolumn{2} {c|} { $R_M$ U. L. $\times 10^{-3}$ } \\
\hline
Belle  &2005  & 1.0 \\
CLEO & 2005 & 7.8 \\
 BABAR & 2004  & 4.2   \\
 FOCUS & 2002  &  1.01 \\
E791 & 1996 & 5.0 \\
 \hline
\end{tabular}
\end{center}
\end{table}

\begin{figure}[bt]
 \vspace{7.5cm}
\includegraphics{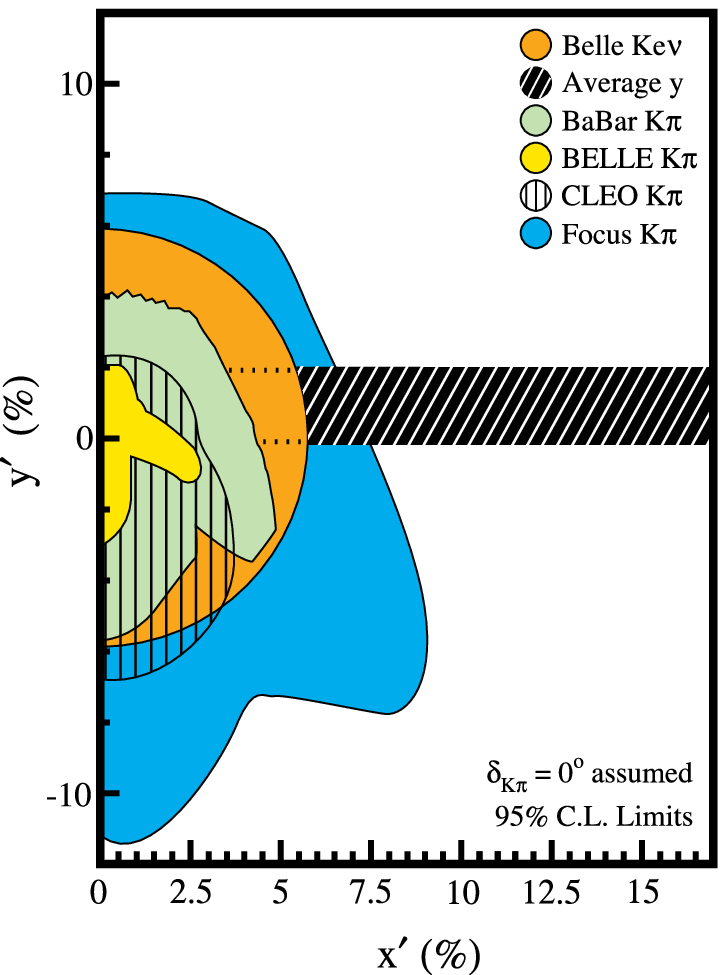}
\caption{\it The status of searches for $D$ meson
mixing at the 95\% C.L. The semicircle $x^2 + y^2$ is the most
restrictive limit from semileptonic decays. For $D^0 \rightarrow
K^+ \pi^-$, $x -y$ contours are shown separately for Belle, BABAR,
CLEO, and FOCUS. For the former three, the limits allow for $CP$
violation in the decay amplitude, the mixing amplitude, and the
interference between these two processes. To place $y$ $\delta_{ k
\pi}=0$ is assumed. The world average allowed range for $y$ is the
horizontal band. If $\delta_{ k \pi} \ne 0$ the allowed $y$
region would rotate clockwise about the origin by an angle
$\delta_{ k \pi}$.}
\label{figure:master-mix}
\end{figure}

Mixing, and $CP$-violation in mixing, can be searched for in a
variety of ways. Measurements of $y$ are summarized in
Table~\ref{table:mix-y} and are reviewed
in~\cite{Annrev}~\cite{Bianco}~\cite{Asner}. The world average is
\begin{equation}
y_{CP} = (0.9 \pm 0.4)\%
\end{equation}
In the limit of $CP$-conservation $y{_{CP}}=y$. The 95\% C.L.
range of $y$ is the horizontal band in
Figure~\ref{figure:master-mix}.


Searching for $D$ mixing in semileptonic decays is straightforward
as there is an unambiguous signal that mixing has occurred:
\begin{equation}
D^{*+} \rightarrow D^0 \pi^+_{\rm tag}, D^0 \rightarrow K^- e^+
\nu_e ~{\rm unmixed}
\end{equation}
\begin{equation}
D^{*+} \rightarrow D^0 \pi^+_{\rm tag}, D^0 \rightarrow
\overline{D^0} \rightarrow K^+ e^- \overline{\nu_e} ~{\rm mixed}
\end{equation}
The flavor of the $D$ meson at birth is tagged by the sign of the
pion from the $D^{*}$, the flavor at decay is tagged by the sign
of the lepton. The time evolution of a neutral $D$ meson depends
on the type of state into which it decays, and it is particularly
straightforward for semileptonic final states.
\begin{equation}
\Gamma_{\rm unmix} \propto  e^{-t/\tau}
\end{equation}
\begin{equation}
\Gamma_{\rm mix} \cong  e^{-t/ \tau} (t/ \tau)^2 {1 \over 4} (
x^2+y^2)
\end{equation}
where $t$ is the proper time of the $D^0$ decay, and the
approximation is valid in the limit of small mixing rates. The
time integrated mixing rate relative to the unmixed rate is
\begin{equation}
R_{\rm mix} = { 1 \over 2} (x^2 + y^2)
\end{equation}
The rate depends quadratically on $x$ and $y$ and does not provide
a way to differentiate between them.
Table~\ref{table:mix-SL} is a compilation of results. The 95\% CL
limit on $R_{\rm mix}$, a circular region centered on $(x=0,y=0)$
is displayed in Figure~\ref{figure:master-mix}.

\begin{table}[tbp]
\centering
\caption{Mixing searches using $D^0 \rightarrow K^+ \pi^-$.
Comparison of the 95\%~C.L. limits in per cent for the fit output
parameters when $CP$ conservation is assumed in the fit. The FOCUS
entries are one dimensional limits. }
\vskip 0.1 in
\centerline{
\begin{tabular}{|cc c c | }
\hline & $R_D$ [$\times 10^{-3}$] & $y^\prime$ [\%] & $x^{\prime 2}/2$ [\%]  \\
\hline 2006 Belle~\cite{BELLE_MIX_KPI} & $3.77 \pm 0.08 \pm 0.05$  & (-2.8, 2.1) & $<0.036$\\
2005 FOCUS~\cite{FOCUS_KPI_WS}& $4.29 \pm 0.63 \pm 0.28$ & (-11.0, 6.6) & $<0.385$ \\
2003  BABAR~\cite{BABAR_MIX_KPI}    & $3.57 \pm 0.22 \pm 0.27$         & (-5.6, 3.9) & $<0.11$ \\
2000 CLEO~\cite{CLEO_MIX_KPI}& $3.32^{+0.63}_{-0.65}\pm0.40$ & (-5.8, 1.0) & $<0.041$\\
 \hline
\end{tabular}}
\label{table:x-y-comp}
\end{table}

Another way to search for $D$ mixing is in the hadronic decay $D^0
\rightarrow K^- \pi^+$. This method is sensitive to a linear
function of $x^2$ and $y$, and can differentiate between them. The
most restrictive  mixing constraints come from this mode. The
unmixed signal is the Cabibbo favored (CF) $D^0 \rightarrow K^-
\pi^+$. The mixed signal is $D^0 \rightarrow \bar{D^0} \rightarrow
K^+ \pi^-$ but it has a background from doubly Cabibbo suppressed
(DCS) decays $D^0 \rightarrow K^+ \pi^-$. Interference between the
CF and DCS decays,  which is linear in $y^ \prime$, gives rise to
the power of the method. The proper decay time distribution is fit
to distinguish between DCS and the mixing signal. For $|x|, |y|
\ll 1$ and negligible $CP$-violation, the decay time distribution
for $D^0 \rightarrow K^+ \pi^-$ is
\begin{equation}
{dN \over dt} = [ R_D + \sqrt{R_D} y^\prime \Gamma t + 1/4
(x^{\prime 2} + y^{\prime 2} ) (\Gamma t)^2 ]e^{-\Gamma t}
\label{equation:kpi}
\end{equation}
where $R_D$ is the ratio of DCS to CF decay rates.  In principle,
there is a strong phase difference, $\delta_{K \pi}$, between the
CF and DCS amplitudes which rotates $x$ and $y$ to $x^\prime$ and
$y^\prime$. To search for $CP$-violation one determines $R_D,
~x^\prime {\rm ~ and~} y^\prime$  separately for $D^0$ and
$\bar{D^0}$. Most recent analyses have been made both with and
without requiring $CP$ conservation.
Table~\ref{table:x-y-comp} is a compilation of results.
Figure~\ref{figure:master-mix} shows: (a) That the Belle recent
analysis is an impressive step forward in sensitivity. (b) There
remains no statistically significant evidence for $D$ meson mixing
although the situation is becoming increasingly tantalizing.


At a charm factory as $\psi(3770) \rightarrow D \overline{D}$ and
$ C=-1$ quantum coherence guarantees that the mixing signature $
D^0 \rightarrow K^- \pi^+ ~,~ \overline{D^0}  \to D^0 \to K^-
\pi^+$, cannot be mimicked by one $D$ undergoing a  DCS decay.
Combining with semileptonic decays to increase sensitivity, a
0.75/fb (15/fb) sample reaches $x < 1.7\% ~~ ( x < 0.4\%)$. A more
sophisticated approach: The Quantum Correlated Analysis (TQCA)
makes a combined fit to single and double flavor and $CP$ tag
yields, which are a function of ${\cal B}_i, x^2 , y, \delta_i$.
TQCA is estimated to achieve a sensitivity for 0.75/fb (10/fb,
1,000/fb) of $x < 2.4\%$  $(x<1.3\%, x< 0.1\%)$ and $y < 1.2\%$
$(y< 0.3\%, y< 0.03\%)$. Purohit showed at this workshop that a B
Factory with 10/ab in a  $D(t) \to K^- \pi^+$ analysis reaches
$x<1\%$ and Nakada showed that at LHC-b in one year an analysis of
$D(t) \rightarrow K^0 \pi \pi$ reaches a sensitivity of $x
<0.4\%$.

\subsection{Measurement of the hadronic phase}

At the $\psi(3770)$ if a  $D^0$ is observed to decay to a $CP$
eigenstate which is $CP$ even: then in the limit of $CP$
conservation, the state recoiling against the tag has a definite
$CP$ as well and it must be of opposite sign, in this case $CP$
odd. Consider the situation where the second $D$ decays to a
flavor mode:
\begin{eqnarray}
\sqrt{2} A( D_{CP\pm} \rightarrow K^- \pi^+) \nonumber \\
 = A(D^0
\rightarrow K^- \pi^+) \pm A(\overline{D^0} \rightarrow K^- \pi^+)
\end{eqnarray}
which defines two triangles from which $\cos \delta_{K \pi}$ can
be determined. Determining $ \delta_{K \pi}$ is necessary to
rotate $x^\prime$ and $y^\prime$ measured in $D(t) \to K \pi^-$ to
$x$ and $y$. The method is limited by the number of $CP$ tags, but
can be extended to many modes simultaneously in the TQCA where the
sensitivity for 0.75/fb (10 /fb) is $\cos \delta_{K \pi} \pm 0.13$
$(\pm 0.05)$.

\subsection{Charm contributions to $\phi_3 / \Gamma$}

The phase of $V_{ub}$, $\phi_3 / \Gamma$ can be determined by the
interference between $ b \to u$ and $ b \to c$ decays where the
$D$ decays to a $CP$ eigenstate or a flavor mode. In the first the
$D$ mixing parameters are needed and in the second knowledge of
$\cos \delta_{K \pi}$. Both methods require very large integrated
luminosity. A third method, the Dalitz method, is currently the
most accessible method experimentally. Here $B \to D K^+ , D \to
K_S^0 \pi \pi$ With this approach the B factories have measured $
\phi_3 = 68 \pm14 \pm 13 \pm 11 ^\circ$~\cite{Belle-gamma}, and
$\gamma = 67 \pm 28 \pm 13 \pm 11$~\cite{BABAR-gamma}, where the
third uncertainty is from the $D$ decay model and can be reduced
by analyzing $CP$ tagged Dalitz plots for $D \to K_S^0 \pi \pi$ at
charm factories. A study by Bondar~\cite{Bondar-gamma} estimates
statistical uncertainty on $\phi_3 / \gamma$ at a B-Factory from
$B \to DK^+$ to be ($\pm 6^\circ$ for 1/ab and $ \pm 2^\circ$ for
10/ab. The integrated luminosity needed to provide the number of
$CP$ tagged $D \to K_S^0 \pi \pi$ to match the statistical
uncertainty from the super B factory for $\pm 6^\circ$  is 0.75/fb
and  $\pm 2^\circ$ for 10/fb. The latter is a good match to the
capabilities of BES III. CLEO-c sensitivity (281/pb) is consistent
with Bondar's prediction

\subsection{Charm CP Violation}

Three types of $CP$ violation are possible. (1) $CP$ violation in
the $D^0  - \bar{D^0}$ mixing matrix. As $D$ mixing is very small,
$CP$-violation in $D$ mixing, commonly parameterized by $A_{\rm
m}$, is negligible both in the SM and many of its extensions.
Experiments are not yet statistically sensitive to it, and so we
will not consider it. (2) $CP$ violation in the interference
between mixing and decay. It is time dependent, since mixing is
involved but it is also small since $D$ mixing is suppressed. It
is a good place to search for new physics, but experiment is only
now becoming sensitive enough. (3) Direct $CP$ violation. This
occurs when the absolute value of the $D$ decay amplitude to a
final state $f$ is not equal to the $CP$-conjugate amplitude


For direct $CP$-violation to occur, two amplitudes with different
weak phases and different strong phases must contribute to the
decay process. The expression for the $CP$ asymmetry $A_{CP}$ is
\begin{eqnarray}
A_{\rm CP}  =  { { \Gamma(D^0 \rightarrow f) - \Gamma (\bar{D^0}
\rightarrow f) } \over { \Gamma (D^0 \rightarrow f) + \Gamma
(\bar{D^0} \rightarrow f) } }  \nonumber  \\
= { { 2 {\it Im} A_1A_2 \sin(\delta_1-\delta_2)} \over {|A_1|^2 +
|A_2|^2 + 2 Re A_1 A_2^\ast \cos(\delta_1 -\delta_2)}}
\end{eqnarray}
where $A_i$, $\delta_1$ and
$\delta_2$ are the moduli of the amplitudes, the weak phase
difference and the strong phase differences, respectively.

In the SM, direct $CP$ violation in the $D$ meson system occurs
for singly Cabibbo suppressed decays such as $D^0 \rightarrow
\pi^+\pi^- / K^+ K^-/K^+K^- \pi^+$, because for these decays there
are several candidates for the second weak amplitude including
penguin graphs, WA diagrams for $D_s^+$ decays, and channels with
a $K_S^0$~\cite{Tony}.

\begin{figure}[btp]
 \vspace{5.5cm}
\includegraphics{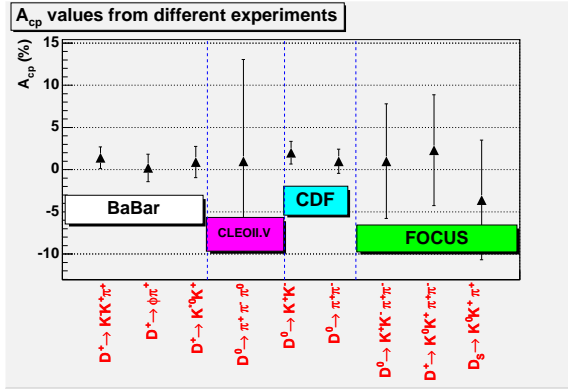}
\caption{\it Selected searches for direct $CP$ violation in $D$
decay.} \label{figure:acp}
\end{figure}

Predictions for $A_{CP}$ are difficult due to the unknown strong
phase. In the SM $A_{CP} < 10^{-3}$. New physics can produce
$A_{CP} \approx 1\%$. However, if an asymmetry at the 1\% level
was observed, one could not rule out a hadronic enhancement of the
SM. Therefore it is necessary to analyze many channels to
elucidate the source of CP violation. Selected measurements of
$A_{\rm CP}$ are shown in Figure~\ref{figure:acp}. Sensitivity
approaches 1\%.


At the $\psi(3770)$ CP violating asymmetries can be measured by
searching for events with two CP odd or two CP even final states.
For $D \to K^+ K^-$ charm factory sensitivity for (0.75/fb ,
10/fb, 1,000/fb) is $(A_{CP} < 0.08, <4 \times 10^{-3}, <6 \times
10^{-5})$ at 90\% C.L. Nakada has shown that in one year at LHC-b
the sensitivity is $A_{CP}< 1.4 \times 10^{-4}$. An alternative is
to search for $CP$ violation in $D \to f , D_{CP} \to ~{\rm flavor
~mode}$. The sensitivities at charm factories are $( A_{CP} <
0.025, < 6 \times 10^{-3}, <7 \times 10^{-4})$ at 90\% C.L.
respectively.

Alternative search strategies include Dalitz plot analyses that
are particularly sensitive since they probe $CP$-violating phases
in the amplitude rather than in the rate, are beginning to be
attempted. These can be performed at charm threshold exploiting
the quantum coherence and at higher energies.



\subsection{Rare Decays}
\begin{figure}[btp]
 \vspace{6.0cm}
\includegraphics{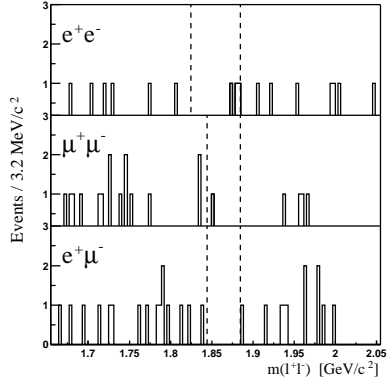}
 \caption{\it The di-lepton invariant mass distribution
  for $D^0 \rightarrow e^+ e^-$, $D^0 \rightarrow \mu^+ \mu^-$ and $
  D^0 \rightarrow e^\pm \mu^\mp $. The dashed lines indicate the
  signal mass region.  } \label{FCNC-search}
\end{figure}

In the SM flavor changing neutral currents are suppressed by the
GIM mechanism. The dilepton decay proceeds by penguin annihilation
or a box diagram.
SM expected branching ratios are ${\cal B}(D^0 \rightarrow e^+
e^-) \sim 10^{-23}$, ${\cal B}(D^0 \rightarrow \mu^+ \mu^-) \sim 3
\times 10^{-13}$. The lepton flavor violating mode $ D^0
\rightarrow e^\pm \mu^\mp$ is strictly forbidden. New physics may
enhance these processes. For example $R$-parity violating SUSY
predicts ${\cal B}(D^0 \rightarrow e^+ e^-) \leq 10^{-10}$, ${\cal
B}(D^0 \rightarrow \mu^+ \mu^-) \leq 10^{-6}$ and  $ D^0
\rightarrow e^\pm \mu^\mp \leq 10^{-6}$~\cite{Gustavo}.
The result of a BABAR search which significantly improved upon
previous upper limits~\cite{BABAR_rare}, is shown in
Figure~\ref{FCNC-search} and Table~\ref{table:rare-BABAR}
respectively.  BES III will reach a sensitivity of $ {\rm few~}
\times 10^{-7}$ a super flavour factory at 10 GeV with 50/ab will
achieve $ {\rm a few~} \times 10^{-9}$ and if operated at the
$\psi(3770)$ also  $ {\rm a few~} \times 10^{-9}$. However the 10
GeV measurement is likely to be compromised by large backgrounds
while the $\psi(3770)$ measurement will have little background and
so the latter will be far superior, although still four orders of
magnitude above the SM rate.

If new physics is present in rare $D$ decays it is likely to be
more experimentally accessible in the modes $ D \rightarrow X
\ell^+ \ell^-$.  In the SM the ${\cal B}(D^+ \rightarrow \pi e^+
e^- ) = 2.0 \times 10^{-6}$. In R-parity violating SUSY the
integrated rate increases by only 20\%, however the differential
dilepton mass distribution is significantly modified compared to
the SM at low and high dilepton masses well away from the
$\rho/\omega/\phi$ SM contributions. Several experiments have
recently made searches, see Table~\ref{table:rare-BABAR}. If $D^+
\to \pi e^+ e^-$ is at the SM level, only 1 evt/fb will be
produced at the $\psi(3770)$ implying BESIII will observe 24
events (including a factor of two for $D^+ \to \pi \mu^+ \mu^-$.)
While this sounds modest, just imagine if these event are
clustered at low or high dilepton mass well away from SM
contributions, it would be clear evidence for new physics!
\begin{table}[tbp]
\centering
\caption{Selected recent searches for rare $D$ decays.}
\vskip 0.1 in
 \label{table:rare-BABAR}
\begin{tabular}{@{}|lll|}
\hline
Mode                      & \multicolumn{2} {c|} { Upper Limit $\times 10^{-6}$ } \\
\hline
$ e^+e^-$ & BABAR &1.2  \\
$\mu^+ \mu^- $  & BABAR & 1.3   \\
$ e^\pm \mu^\mp $ & BABAR       &  0.81   \\
$\pi e^+ e^-$ & CLEO-c & 7.4 \\
$K e^+ e^-$ & BABAR & 3.6 \\
\hline
\end{tabular}
\end{table}

In summary the  experimental sensitivity for both $D^0$ and $D^+$
rare decays is in the range $10^{-5} - 10^{-6}$. For some modes,
notably $D \rightarrow \pi^+ \ell^+ \ell^-$, measurements are
beginning to confront models of new physics. In other cases,
measurements are far above the SM prediction. The outlook for
rare charm decays is promising. CDF, the B Factories, the charm
factories, ATLAS/CMS and LHC-b will all contribute. For selected
projections see~\cite{Annrev}.


\section{SUMMARY}

New physics searches in the charm sector involving mixing,
$CP-$violation and rare decays have become considerably more
sensitive in the past several years, however, all results are
null.

In charm's role as a natural testing ground for QCD techniques,
there has been solid progress. Data at the $\psi(3770)$ from BESII
and CLEO-c, and later BESIII, is finally producing a new era of
precision absolute charm branching ratios.
This is well-matched to developments in theory, especially the
lattice, which has a goal to calculate to a few percent precision
in the $D, B, \Upsilon, {\rm ~and~} \psi$ systems. CLEO-c, and
later BES III, will provide few per cent precision tests of
lattice calculations in the $D$ system and in heavy onia, which
will quantify the accuracy for the application of LQCD to the $B$
system. If all goes to plan,  BABAR, Belle, CDF, D0, CMS, ATLAS,
and LHC-b data, in combination with LQCD will produce a few per
cent determinations of $|\vub|, |\vcb|, |\vtd|,$ and $|\vts|$
thereby maximizing the sensitivity of the flavor physics program
to new physics beyond the SM this decade and aid understanding
beyond the SM physics at the LHC in the coming decade.


\section{Acknowledgements}
I thank my colleagues on BABAR, Belle, BES II, CDF, CLEO, D0,
FOCUS and LHC-b for many valuable discussions. Bo Xin is thanked
for technical assistance. I am particularly grateful to Stefano
Bianco for organizing a superb meeting in Frascati and for his
patience while I completed this manuscript.

\end{document}